\documentclass[%
 aps,
 amsmath,amssymb,
 preprint,%
 onecolumn,
 longbibliography, superscriptaddress
]{revtex4-1}
\usepackage{multirow}
\usepackage{hyperref}
\usepackage{graphicx}% Include figure files
\usepackage{dcolumn}% Align table columns on decimal point
\usepackage{bm}% bold math
\usepackage{subfig}
\begin{document}

%\preprint{AIP/123-QED}

\title{A network partition method for solving large-scale complex nonlinear processes}

\author{Shucheng Pan}
\email{shucheng.pan@tum.de}
\author{Jianhang Wang}
\email{jianhang.wang@tum.de}
\author{Xiangyu Hu}%
\email{xiangyu.hu@tum.de}
\author{Nikolaus A. Adams}
\email{nikolaus.admas@tum.de}
\affiliation{Department of Mechanical Engineering, Technical University of Munich, 85748 Garching, Germany}

\date{\today}% It is always \today, today,

%
\begin{abstract}
A numerical framework based on network partition and operator splitting is developed to solve 
nonlinear differential equations of large-scale dynamic processes encountered in physics, chemistry and biology. Under the assumption that those dynamic processes can be characterized by sparse networks, we minimize the number of splitting for constructing subproblems by network partition. Then the numerical simulation of the original system is simplified by solving a small number of subproblems, with each containing uncorrelated elementary processes. In this way, numerical difficulties of conventional methods encountered in large-scale systems such as numerical instability, negative solutions, and convergence issue are avoided. In addition, parallel simulations for each subproblem can be achieved, which is beneficial for large-scale systems. Examples with complex underlying nonlinear processes, including chemical reactions and reaction-diffusion on networks, demonstrate that this method generates convergent solution in a efficient and robust way.
\end{abstract}

\maketitle

\section{Introduction \label{sec:intro}}
Solving nonlinear dynamic processes which are ubiquitous in a broad range of physical, biological, and social systems is numerically challenging when the scale and complexity become large. The high accuracy, efficiency and robustness of classic numerical methods exhibited in small-scale simple nonlinear processes are deteriorated when applied to large complex systems, of which chemical reaction is a canonical example where multiple reacting species are governed by complex kinematics with a large number of reactions. The vastly disparate timescale of reactions leads to high stiffness of the system \cite{oran2005numerical}. For instance, a typical chemical reaction mechanism in combustion consists of various timescales spaning from e.g. $10^{-4}$s to $10^{-12}$s \cite{lu2009toward, gou2010dynamic}. This prohibits the use of explicit numerical methods such as Euler scheme and Rungu-Kutta schemes, as the timestep requred by stability condition is about $10^{-12}$ which is several order smaller than the timescale of turbulent mixing \cite{gou2010dynamic}. Thus in the combustion community, implicit methods such as ref. \cite{brown1989vode} are commonly employed to handle the reaction terms with large timesteps due to the numerical roubstness. However those methods are only efficient for simple kinetic mechanisms as the operation complexity scales cubically with the the size of chemical kinetics \cite{lu2009toward}. For example, even for a combustion problem with simple methane chemistry (35 species and 217 reactions), more than 90\% of the overall computational time is spent on chemistry calculations \cite{schwer2003adaptive}. Thus numerical simulations of practical combustion problems with detailed chemical mechanism remain to be challenging due to the high stiffness and large number of species and reactions (in the order of $\mathcal{O}(10^3)$) \cite{lu2009toward}, espeically in three dimensions \cite{diegelmann2017three}. 
Using quasi-steady-state (QSS) approximation to handle fast reactions eliminate the stiffness of the system but this treatment the violates mass conservation properties and requires ad hoc determination of QSS species \cite{lu2009toward}.

Another example is the dynamic processes, say reaction-diffusion, on a complex network \cite{colizza2007reaction, nakao2010turing, asllani2014theory} which have been widely used to represent behaviours in diseases spreading \cite{balcan2009multiscale}, protein-protein interaction \cite{rain2001protein}, and regulation of gene expression \cite{karlebach2008modelling}. Unlike the nonlinear processes under the homogeneous assumption of the substrate, which can be easily solved in regular lattice, the high heterogeneity and large-scale of complex networks present challenges for numerical solving the underlying nonlinear dynamics \cite{nakao2010turing}. In this case, the implicit numerical methods are difficult to be applied due to the heterogeneous structures and thus timestep is chosen to be small when the diffusion rate is large, leading to large computational cost.

Instead of solving the complicated system directly, operator splitting, such as Lie splitting scheme \cite{lie1888theorie}, decomposes the original problem into smaller subproblems based on different mathematical and physical properties, with each individual easily solved by dedicated methods. Compared to classic non-splitting numerical methods, the operator splitting method is more efficiency and easier to converge. 
It requires less amounts of memory and offers flexibility to select suitable discretized schemes as well \cite{glowinski2017splitting}. If the subproblems are solved suitably, usually by using local implicit methods, numerically solving the original problems by the operator splitting becomes unconditionally stable. Many well-established numerical methods have been developed based on the operator splitting concept to solve different physical and mathematical problems \cite{glowinski2017splitting}. One example is the projection method \cite{chorin1997numerical} which separately solves the velocity and the pressure field of the incompressible Navier-Stokes equations. Another one is the split Bregman method \cite{goldstein2009split} which is developed to solve optimization problems in imaging processing, compressive sensing, and machine learning.

Canonical applications of splitting methods include the splitting reaction terms from convection terms for reacting flows \cite{leveque1990study} and splitting of reaction and diffusion terms for reaction-diffusion equations \cite{descombes2001convergence}. For a reacting system with a small number of subprobelms (operators) $N$, the splitting can be applied to every elementary reactions and the results indicate that this treatment is unconditional stable and preserves conservation property \cite{nguyen2009mass}. However, in many complex system, $N$ is large, implying that directly applying existing schemes requires a large number of splitting \cite{nguyen2009mass, litvinov2010splitting}. This leads to large splitting errors and difficulty to impose parallelization as the $N$ subproblems are solved sequentially. Many large-scale complex dynamic systems exhibit network structures whose nodes represent different spatial elements or physical terms. The topology of those networks, although very complex, is highly sparse in the sense that the number of interactions between different nodes in the network is small. For example, the species or elementary reactions are sparsely coupled in large chemical reacting systems \cite{deuflhard1986efficient}. This motivates the development of an efficient method that ultizes sparsity of network structures to solve large-scale nonlinear processes with sparse network structures. The key idea is applying the network partition, which previously is used for e.g. detecting community structure \cite{newman2012communities} and serves to ensure only uncorrelated nodes being clustered into the same subset here, on numerical solving large-scale time-dependant dynamic system.
We expect the proposed method to be efficient, convergent, and numerical stable for solving complex nonlinear dynamics on a extremely large system. After dwelling on the method and its main features, a variety of numerical examples are tested to demonstrate the main features.

\section{The network partition for a complex dynamic system} \label{sec:method}
Given a large-scale complex system with time-dependent nonlinear dynamics, we first partition the network based on some chosen principals to achieve a small number of subprolems, each is a subset of decoupled processes that can be solved by classic analytical or numerical methods. This is carried out by three steps: (i) abstract the system with a network $\Gamma$ which usually is sparse and (ii) translate $\Gamma$ to a graph $G$ by a specific mapping $f: \Gamma \rightarrow G$ and (iii) partition $G$ and package the nodes to generate a subproblem. Use operator splitting methods, the subproblems are solved sequentially, that is, the solution of previous subproblem is the input of the next one. 
Here we use an illustrative example corresponding to a chemical reaction in Fig.\ref{fig:network} to describe the details.

\subsection{The network of a nonlinear complex process} \label{sec:net}
\begin{figure}[h]
\centering
\hspace{0.3cm} \includegraphics[width=0.7\linewidth]{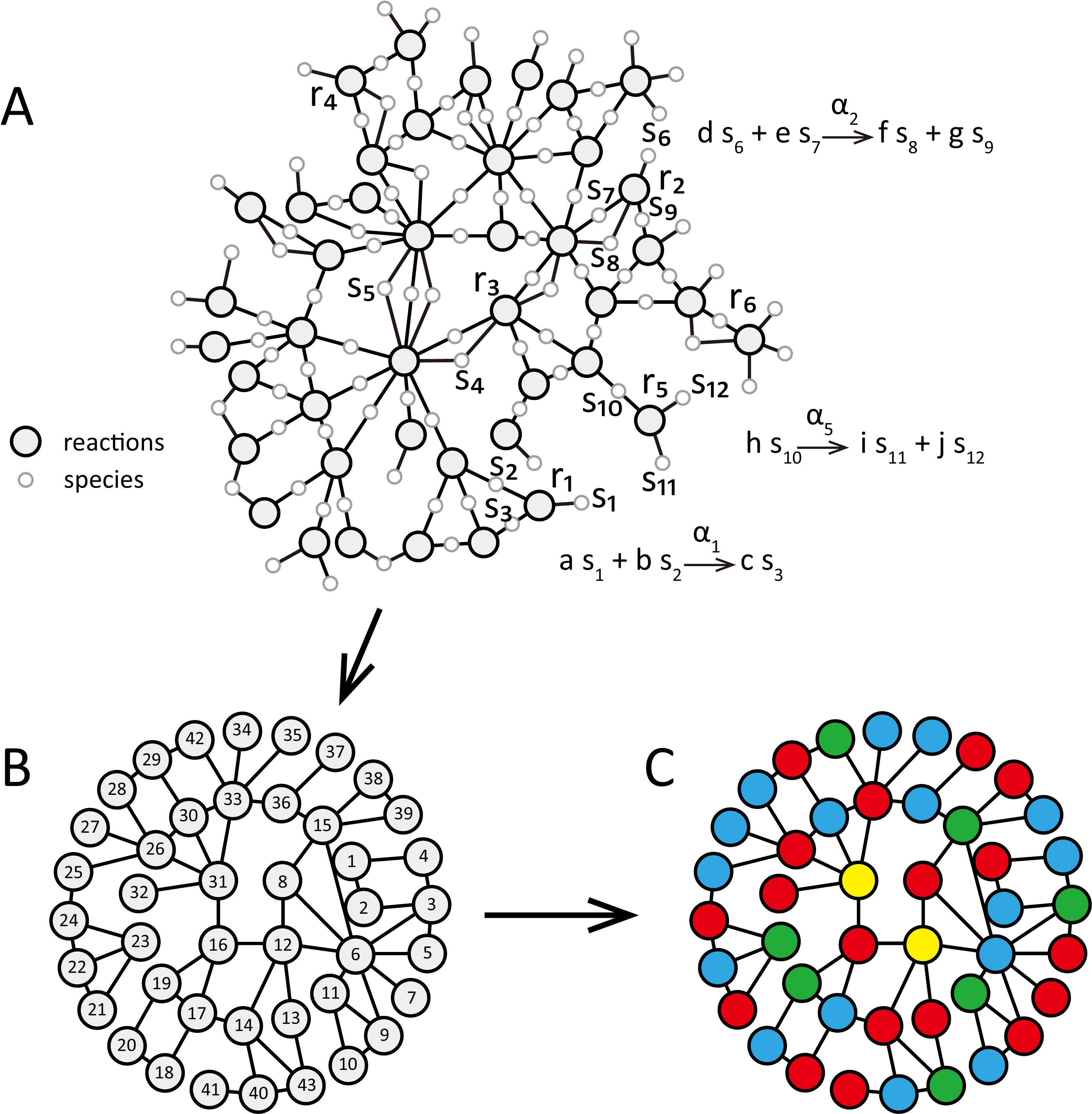}
\caption{An illustrative example of network partition method for chemical reactions. A reaction system involving $42$ channels can be represented by a network $\Gamma$ in (A) and is mapped to a planar graph $G$ (B), where we consider a primitive mapping $f: \Gamma \rightarrow G$ that an edge $e(x,y)$ exists when two reactions $x$ and $y$ has common species. After partition $G$ by graph coloring algorithm with $4$ colors (C), we package the reactions with the same color, which are evolved simultaneously without any interaction.}
\label{fig:network}
\end{figure}

As shown in Fig.\ref{fig:network}A, a chemical reaction system with $43$ channels exhibits a network structure. In this network, each node (reaction) has multiple interactions with others by affecting the a number of reacting species. We can find a mapping $f: \Gamma \rightarrow G$ that convert the network to a graph $G=(V, E)$, where $V$ and $E$ are the set of nodes (elements) and the set of edges (pairs), respectively. For instance, one can use the definition in Fig.\ref{fig:network}B where multiple interactions between the same pair of reactions, $(x,y)$, collapse to one edge $e(x,y)$, i.e., the element $a(x,y)$ of the adjacency matrix $\mathbf{A}(G)$ is $1$ or $0$, depending on whether the reactions $x$ ad $y$ evolve common species. The degree of each node $x$, $d(x) = \sum_{y\in V} a(x,y)$, is highly heterogeneous and the distribution of edges shows sparsity, i.e., the number of edges $n(E)$ is considerably smaller than of the complete graph. 

Besides, the node in $G$ can be a group of coupled elementary processes which is solvable in the sense that it has analytical solutions or is easy to be numerically solved. For example, multiple reactions in Fig.\ref{fig:network} interacted by common product species have analytical solutions and thus can be considered as a single node in $G$, leading to a mapping different $f^{*}$. Or one define a group of reactions as a node if it is efficient to be solved by locally applying implicit time-integration methods. Then the network partition is performed to ensure there is no interaction between those small structures if they are grouped into the same subset.
\subsection{Operator splitting and network partition methods} \label{sec:split}
Large-scale systems are too difficult to solve directly, either due to high computational cost or numerical instabilities. Instead, we apply the operator splitting method which separates the original system into a number of small-scale subsystems, with each one being easy to be numerically solved, and thus offering convenience for solving large-scale nonlinear systems. For instance, the reaction system in Fig.\ref{fig:network} can be solved by e.g. Lie splitting method, i.e., the $N$ elementary reactions are solved individually. In this way the results are positive and conservative without time-step constraints and omits costly matrix operations \cite{nguyen2009mass}. 

When $N$ is large, the large number of splitting generates significant splitting errors which prevent the use of large time-steps. Another issue is that the operator splitting method requires a sequential updating of all $N$ subproblems. This causality leads to difficulty for parallization which is usually required for solving large-scale problems. To reduce the number of splitting and alleviate the causality of operator splitting methods, we propose a numerical framework that utilizes the sparsity of the network structure. Usually in most of large-scale systems the directly interacting elements are sparse and the elements that are not direct interacted can be solved simultaneously without numerical difficulty. This inspires us to split the original system into $K$ ($K<N$) subsystems where their elements have no direct interaction and solves the dynamics of all elements inside a certain subsystem simultaneously. 
This leads to a network partition problem: given a graph $G(V,E)$ after mapping from a physical system, say chemical reaction, one cluster its nodes by a partition $S=\bigcup_{k=1}^{K}S_k$ that satisfies
\begin{equation} \label{coloring}
S = \underset{s}{\text{arg min}}\, K_s \quad \text{subject to} \quad \mathbf{A}_k = \mathbf{0}, 
\end{equation}
where $k$ indexes the subsets and $\mathbf{A}_k$ is the adjacency matrix of the $k$-th subset. This partition can be achieved by a graph coloring algorithm. When the network is mapped to a planar graph, one have $K \leqslant 4$. 

After partitioning the network, the elementary processes clustered into the same subset, say $S_k$, generate a subproblem which can be solved in parallel, as the causality updating constraint is avoided in $S_k$. Consider $N_{k}$ elementary irrelevant reactions belong to the subset $S_{k}$. The $N_{k}$ processes are solved simultaneously and the updated involved species are chosen to be input of $S_{k+1}$.
The entire reacting system is numerically solved by Lie splitting scheme, $\mathbf{X}^{n+1} = \mathbf{E}_1(\Delta t) \circ \mathbf{E}_2(\Delta t) \circ \mathbf{E}_3(\Delta t)$,
if $K=3$, where $\mathbf{X} \in \mathbb{R}^M$ is the concentration vector and $\mathbf{E}_k$ is the time-integration of the subset $S_k$. Note that the network partition is a semi-constructive procedure and $\mathbf{E}$ can be prescribed, say the backward Euler scheme. In this paper, we use analytical solution according the type of every elementary reaction.

The splitting error arising from decoupling treatment of the original couple system can be further reduced by some existing strategies \cite{speth2013balanced}. And the order of accuracy, limited to 1st order due to Lie splitting, can be increased by higher order operator splitting schemes \cite{strang1968construction, descombes2001convergence,thalhammer2008high}. Consider the widely used Strang splitting scheme \cite{strang1968construction}, we can arrange the decoupled subproblems symmetrically and solve them by $\mathbf{X}^{n+1} = \mathbf{E}_1(\Delta t/2) \circ \mathbf{E}_2(\Delta t/2) \circ \mathbf{E}_3(\Delta t) \circ \mathbf{E}_2(\Delta t/2) \circ \mathbf{E}_1(\Delta t/2)$.
The adaptive time-stepping techniques can be applied to improve the accuracy based on local errors estimation. Here we use a similar way with ref.\cite{litvinov2010splitting} to control the local errors. We reduce the time-step if the relative error between two solutions of different level are larger than a tolerance $\varepsilon \Delta t^k$, where $k$ is the order of splitting method and $\varepsilon$ is a small constant. Consider the base time-step is $\Delta t$ and we perform $n^0_s$ full numerical evaluations at the coarsest level. If the refinement level is $\ell$, the time-step can be reduced to $\Delta t/(2^{\ell} n^0_s)$. 

And there exists a large number of partition strategies for large-scale system. 
Indeed, the optimal partition is the one that minimizes the splitting error, say $\frac{\Delta t^2}{2}[\mathbf{M}_1,\mathbf{M}_2]\mathbf{X^0}$ for $K=2$, where $[\cdot]$ is the commutator and $\mathbf{M}$ is the operator for each subproblem. However this strategy requires applying optimization on-the-fly, which is costly for large $N$. Alternatively, we can find a partition that the most relevant nodes are clustered into the same subset. One can relate the network with a Markov chain and define a metric called ``diffusion map'' which measures the correlation of different nodes \cite{lafon2006diffusion}. For a reacting system, a weight matrix that measures the pairwise interaction strength is defined as $w(x,y) = \sum_{z} \max (\alpha(x) \vert \nu(x,z) + \mu(x,z) \vert, \alpha(y) \vert \nu(y,z) + \mu(y,z) \vert)$, which assembles the connectivity measure in some reaction mechanism reduction methods \cite{lu2005directed}, where $z$ labels all common species of reactions $x$ and $y$, $\alpha$ is the reaction rate coefficient, $\nu(x,z)$ is the stoichiometric coefficient of reactant $z$ in reaction $x$, and $\mu(x,z)$ is the stoichiometric coefficient of product $z$ in reaction $x$. Then the transition matrix of the corresponding Markov chain is determined by $p(x,y) = w(x,y)/\sum_{z \in G} w(x,z)$. And let $\lambda_l$, $\psi_l$ and $\phi_l$ be the eigenvalue, normalized right and left eigenvectors of $P$ with $0\leqslant l \leqslant N-1$. The diffusion map $\Psi_{t}(x) = [\lambda_1^t \psi_{1}(x), \,\, \lambda_2^t \psi_{2}(x), \,\, \cdots , \,\, \lambda_q^t \psi_{q}(x)]^{\text{T}}$ is introduced in ref.\cite{lafon2006diffusion} so that one can determine the connectivity of different nodes by measuring their distance in diffusion coordinates, $D^2_{t}(x,y) = \Vert \Psi_{t}(x) - \Psi_{t}(y)\Vert^2$. Then nodes are clustered into subsets depending on their diffusion distance to the geometric centroids of subsets $c(S_k)=\sum_{x \in S_k} \frac{\phi_{0}(x)}{\bar{\phi}_0(S_k)}\Psi_t(x)$, where $\bar{\phi}_0(S_k) = \sum_{x\in S_k} \phi_0(x)$. This leads to a new partition that minimizing the distortion $\sum_k\sum_{x\in S_k}\phi_0(x)\Vert \Psi_{t}(x) - c(S_k)\Vert^2$. For our case, we follow the clustering algorithm in \cite{lafon2006diffusion} to modify our initial partition, $S^{(0)}=\bigcup_{k=1}^{K}S^{(0)}_k$ which is generated by graph coloring algorithm. Then the partition becomes
\begin{eqnarray} \label{optimal}
S_k^{(p)} &=& \left\lbrace x | k = \underset{l \in [1,K]}{\text{arg\,min}} \Vert \Psi_{t,i} - c(S_l^{(p-1)})\Vert^2 \right\rbrace  \quad \text{subject to} \quad \mathbf{A}_k = \mathbf{0}
\end{eqnarray}
which is solved by constrained k-means algorithm \cite{wagstaff2001constrained} with $p$ indexing the iteration step. Then the most relevant nodes are grouped into the same subset while maintaining the constraint $\mathbf{A}_k = \mathbf{0}$. Although this is not the optimal partition, it outperforms the initial partition generated by coloring algorithm in our test cases.

\subsection{Main features}
The current method has the following advantages when applied to large-scale complex systems:
\begin{enumerate}
\item This method is unconditionally stable, provided the time-integration $\mathbf{E}$ for each subproblem is stable. Thus the chosen time-step can be very large, which offers efficiency for computations of large-scale systems.
\item Negative concentration generation is avoided for systems with large stiffness.
\item The number of splitting, $K$, is small for large-scale complex systems due to the sparsity feature, indicating a reduced splitting error.
\item The method are well suited for parallel simulations as the elements of every subproblem have no direct correlation.
\end{enumerate}
\section{Numerical examples} \label{sec:example}
We start with relatively small problems to analysis the convergence and demonstrate the robustness of the current method. Afterwards, we apply the method to solve large-scale systems to show its efficiency.

\subsection{Chemical reactions} \label{sec:reaction}
Given a well-stirred chemical system with $M$ chemical species which interact through $N$ reactions. If the stochastic effects in this system are neglected and the concentration vector $\mathbf{X} = \{\text{X}_1, \text{X}_2, \cdots, \text{X}_M \}^T$ is assumed to vary continuously in time, the time evolution of the system can be described by the reaction rate equation,
\begin{equation} \label{eq:rre}
\frac{d \text{X}_i}{d t} = \sum_{j=1}^N (\nu_{ji} - \mu_{ji}) \text{C}_j(t),
\end{equation}
for $1 \leq i \leq M$, where $\mu_{ji}$ and $\nu_{ji}$ are stoichiometric coefficients of reactant $i$ and product $i$ involved in the reaction $j$, respectively. $\text{C}_j(t)$ is the rate of reaction $j$ and usually determined by the law of mass action, $\text{C}_j(t) = \alpha_j \prod_{l} \text{X}_l(t)^{\mu_{jl}}$, where $\alpha_j$ is the reaction rate coefficient and the index $l$ labels all involved species in the reaction $j$.
Usually the number of involved species and the reactions are enormous, which presents challenges when numerically solve the underlying kinematics. We will test our method through different types of chemical reactions, including the constant rate-coefficient reaction, temperature dependent reaction and stochastic reaction.

%
%%%%%%%%%%%%%%%%%%%%%%%%%%%%%%%%%%%%%%%%%%%%%%%%%%%%%%%%%%%%%%%%%%%%%%%%%%%%%%%%%%%%%%%%%%%%%%%%%%%%%%
\begin{table}[!ht]%[tbhp]
%\centering
\caption{Convergence results of the apoptosis-regulation reaction network \cite{malik2017control} with $n^0_s=2$ and $\ell = 3$ at $t=100$ hours by solving the reaction rate equation (\ref{eq:rre}). The partition, $S^{(0)}$, which is obtained by graph coloring algorithm and listed in Table \hyperref[sec:si]{IV}, is used. The rate of convergence order is listed in parenthesis. The error tolerance is $\epsilon = 0.01 \Delta t$.}
\label{table:conver}
\begin{tabular}{lrrrr}
\hline
						  & $\Delta t = 8.0$s & $4.0$s & $2.0$s & $1.0$s \\
\hline
	  
\multirow{1}{1.5cm}{p53$_p\_$A}  &   $83.60$ & $47.40$($0.8$) & $24.80$($0.9$) & $12.80$($1.0$) \\
\multirow{1}{1.5cm}{p53$_p\_$p300}  &  $2.420$ & $1.284$($0.9$) & $0.658$($1.0$) & $0.334$($1.0$) \\
\multirow{1}{1.5cm}{p53}  &     $0.920$ & $0.485$($0.9$) & $0.245$($1.0$) & $0.123$($1.0$) \\				  
\multirow{1}{1.5cm}{p300}&   $0.222$ & $0.114$($1.0$) & $0.058$($1.0$) & $0.029$($1.0$) \\
\multirow{1}{1.5cm}{HDAC1}  &   $0.119$ & $0.062$($0.9$) & $0.031$($1.0$) & $0.016$($1.0$) \\
\multirow{1}{1.5cm}{Mdm2} &    $0.077$ & $0.038$($0.9$) & $0.019$($1.0$) & $0.009$($1.0$) \\
\multirow{1}{1.5cm}{p53$_p$}  & $1.8$e-$5$ & $1.8$e-$5$($0.0$) & $1.2$e-$5$($0.6$) & $6.5$e-$6$($0.9$) \\
\multirow{1}{1.5cm}{SMAR1}  &   $1.3$e-$4$ & $6.6$e-$5$($1.0$) & $3.3$e-$5$($1.0$) & $1.7$e-$5$($1.0$) \\  					  				  
\hline
\end{tabular}
\end{table}
%%%%%%%%%%%%%%%%%%%%%%%%%%%%%%%%%%%%%%%%%%%%%%%%%%%%%%%%%%%%%%%%%%%%%%%%%%%%%%%%%%%%%%%%%%%%%%%%%%%%%%

%
\subsubsection{Constant rate-coefficient reaction}
We consider the p53-SMAR1 regulatory biochemical network \cite{malik2017control} whose mechanism, as detailed in Table \hyperref[sec:si]{II}, involves $18$ species (proteins and complexes) and $35$ elementary reactions.
The system is initialized with $\mathbf{X}=\mathbf{0}$ and the constant reaction-rate coefficients are listed in Table \hyperref[sec:si]{II}. Although this case is small, its reaction rates of elementary reaction cross a broad range of magnitudes, $\mathcal{O}(10^{-5}) \sim \mathcal{O}(1)$, indicating high stiffness of this system. We use adaptive time-step mentioned above, where the largest time-step is $\Delta t$ and the the refinement level is $\ell = 3$, i.e., the finest time-step is $\Delta t/8$ if the local errors are larger than the tolerance which is set as $\epsilon = 0.01 \Delta t$ in this case. The elementary reactions are clustered by $K=11$ subsets if we use the mapping in Fig.\ref{fig:network} by considering every reaction is a node in the graph. In Table \ref{table:conver}, we show first-order convergence results by reducing the time-step $\Delta t$ from $8.0$ to $1.0$, for high concentration species, e.g. p53$_p\_$A, and low concentration species, e.g. SMAR1.

\begin{figure}[h]
\centering
\includegraphics[width=1.0\linewidth]{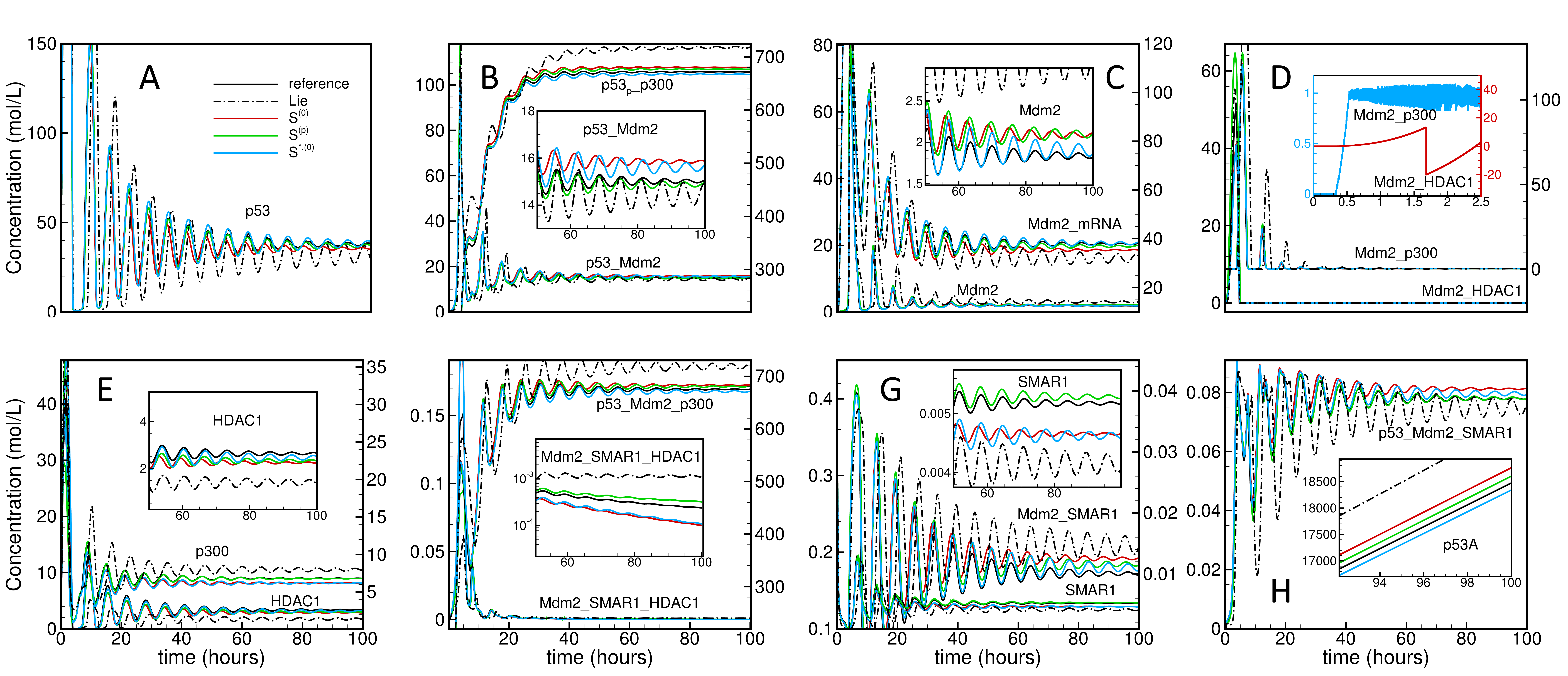}
\caption{Time evolution of the p53-SMAR1 regulatory reaction by the Lie and network partition based splitting methods. The parameters for adaptive time-step are $n^0_s=2$, $\ell = 3$, and $\epsilon = 0.01 \Delta t$. The reference solution is obtained by a small time-step $\Delta t=1s$ while the colored lines indicate solutions with large time-step $\Delta t=32s$ and the dashed lines are the solutions of the Lie splitting method. Three different partition strategies, $S^{(0)}$, $S^{(p)}$, and $S^{*,(0)}$, as listed in Table \hyperref[sec:si]{IV}, are employed. And the insert of Mdm2$\_$p300 and Mdm2$\_$HDAC1 are the results of a 2nd-order Runge-Kutta method.}
\label{fig:apoptosis}
\end{figure}

The ability to use large time-steps demonstrates the robustness of our method. As shown in Fig.\ref{fig:apoptosis}, the time history of the species indicate that the positive $\mathbf{X}$ is preserved for a very large time-step. However, for the 2nd-order Runge-Kutta scheme, numerical instability is observed even we use a small time-step $\Delta t = 1$. Negative concentration of species occur due to spurious solutions on the level of the truncation error, as shown in Fig.\ref{fig:apoptosis}D. We compare the result of our method with the original Lie splitting method. The results of the Lie splitting method show larger errors compared to those of the three solutions obtained by our method. This demonstrates that splitting errors are significantly reduced by applying the network partition, as one motivation of developing the present method.
Then we test different partition strategies listed in Table \hyperref[sec:si]{IV} and compare the results with the reference solution ($\Delta t= 1$). As shown in Fig.\ref{fig:apoptosis}, the results of the optimized partition $S^{(p)}$ in Eq.\ref{optimal} show better agreement with the reference than those of graph coloring based partition $S^{(0)}$ in Eq.\ref{coloring}. If we consider the mapping $f^*$ above, $K$ is reduced to $9$, see $S^{(0),*}$ in Table \hyperref[sec:si]{IV}. And the numerical results are also improved due to a smaller number of splitting.
\subsubsection{Temperature dependent reaction}
This type of chemical reactions are widely encountered in computer simulations of combustion. The reaction rates are determined by the Arrhenius law, $\alpha_j^f = A\, \text{T}^B\, \exp (-E_j/\text{T})$ for forward reactions (odd value $j$) and $\alpha_j^b = \alpha_{j-1}^f / \alpha_j^p$ for backward reactions (even value $j$), where
\begin{align} \label{arrhen}
& \alpha_j^p = \left( \frac{p_0}{\text{RT}}\right)^{\sum_i^M (\nu_{ji} - \mu_{ji})}  \exp  \left( - \sum_i^M W_i\frac{\nu_{ji} - \mu_{ji}}{\text{R}} (\frac{h_i}{\text{T}} - s_i) \right). \nonumber
\end{align}
with $p_0=1$atm. $\text{R}$ is the universal gas constant, $\text{T}$ is the temperature of the system, and $W_i$ is the molecular weight of the species $i$, see Table \hyperref[sec:si]{XI}. The specific internal enthalpy $h_i$ and entropy $s_i$ of the species $i$ are calculated by the equation in \hyperref[sec:si]{Supporting Information}.

We select the hydrogen-oxygen reaction with $9$ species and $23$ reversible reactions (forward and backward). The mechanism and the partition $K=34$ are listed in Table \hyperref[sec:si]{V} and Table \hyperref[sec:si]{VI}, respectively. The initial pressure of the $H_2$-air mixture is $1\,\text{atm}$  and the molar ratio is $2 : 1: 3.76$ for $H_2:O_2:N_2$. Nitrogen is inert and thus treated as a diluent.

\begin{figure}[h]
\centering
\includegraphics[width=0.8\linewidth]{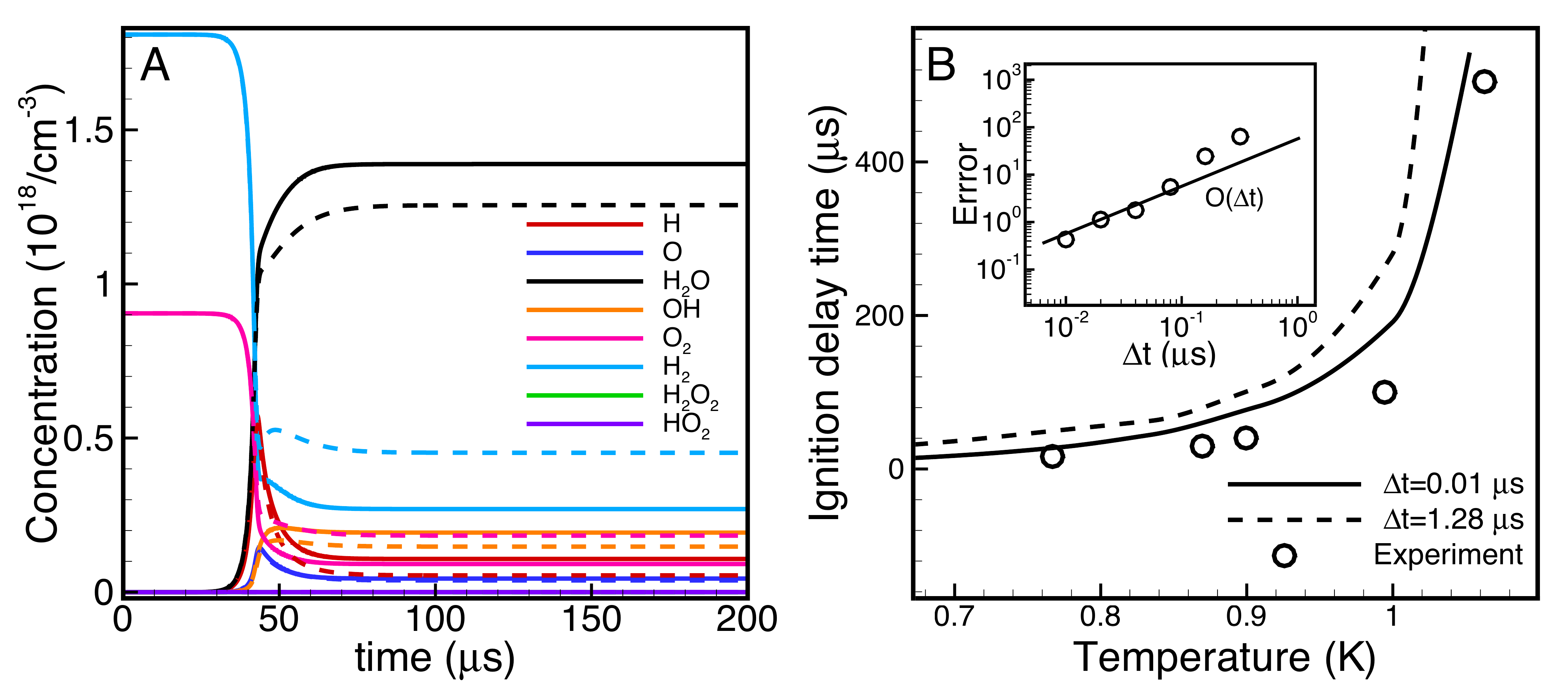}
\caption{Results of the hydrogen-oxygen reaction. The time history of $\mathbf{X}$ is plotted in (A) where a small and large time-step, $\Delta t=0.08 \mu s$ (dashed lines) and $\Delta t=0.01 \mu s$ (solid lines), are chosen. No numerical instabilities or negative concentration are observed. The temperature at $t=200 \mu s$ is measured and its error convergence is shown in (B) with $\Delta t$ increasing from $0.01 \mu s$ to $0.08 \mu s$. The comparison of the ignition delay time between low resolution numerical results ($\Delta t = 0.08 \mu s$), high resolution numerical results ($\Delta t = 0.01 \mu s$), and experimental data \cite{slack1977investigation} is given in (B).}
\label{fig:HO}
\end{figure}

With an initial temperature of $1200 K$, the results calculated by constant time-step ($\ell=1$) $\Delta t=1.28 \mu s$ and $\Delta t=0.01 \mu s$ are shown in Fig.\ref{fig:HO}A. The time evolution of species concentration indicates that numerical instabilities and negative solution are prevented in our method. Before the ignition, the results of very large time-step agree with those of the small time-step. The errors of temperature at $t=200 \mu s$ are measured and the convergence rate shown in Fig.\ref{fig:HO}B is first order, as expected. Then the initial temperature is reduced from $1200 K$ to $950 K$ to verify the computed ignition delay time which is measued by the maximum $d \text{T}/ d t$. As shown in Fig.\ref{fig:HO}B, our numerical results are in good agreement with the experimental data \cite{slack1977investigation}, even with a large time-step. Using 2nd-order Runge-Kutta scheme with $\Delta t=0.01 \mu s$ produce numerical instability and negative concentration of species. Note that the Lie splitting method can produce similar results as this small system is almost fully coupled, leading to similar number of splitting for network partition ($K=34$) and the Lie splitting method ($K=46$).

\subsubsection{Stochastic reaction}
When the molecular populations are relatively small, the dynamic behaviour of the reacting system described by the deterministic differential equation (\ref{eq:rre}) becomes inaccurate. In this case, the stochastic reaction kinetics should be considered \cite{higham2008modeling} and $\mathbf{X} \in \mathbb{N}^M$ becomes the molecular number vector. 
Here we show how to extend our method to stochastic reaction simulation where analytical solutions are more difficult to find than in deterministic problems above. As mentioned above, our method can prescribe time-integration schemes for each subproblem. Then some existing accelerated approximated stochastic methods, such as the $\tau$-leaping method \cite{gillespie2001approximate}, can be employed to simulate a well-stirred system with low molecular number. After partition, the $\tau$-leaping formula,
\begin{equation} \label{eq:tau}
\mathbf{X}^k (t+\tau) = \mathbf{X}^k (t) + \sum_{l=1}^{N_k} (\nu_{l} - \mu_{l}) \mathcal{P}_l \left(a_l(\mathbf{X}^k (t)) \tau\right), \nonumber
\end{equation}
is applied for the subset $S_k$, where $\tau$ is the leap time and $\mathbf{X}^k$ is the molecular number vector of all species belongs to $S_k$. The propensity function $a_l(\mathbf{X}^k (t)) $ \cite{higham2008modeling} of the reaction $l$ is determined by
\begin{equation} \label{eq:propen}
a_l\left(\mathbf{X}^k (t)\right) = \alpha_l \prod_{i} \frac{X_i(t)!}{\mu_{li}!\, \left(X_i(t)-\mu_{li}\right)!}, \nonumber
\end{equation}
where $i$ is the index of all reactants for the reaction $l$. $\mathcal{P}_l\left(a_l(\mathbf{X}^k (t) \tau\right)$ is a Posisson random variable with mean and variance being $a_l(\mathbf{X}^k (t) \tau)$.

For reactants with small molecular number, the unbounded $\mathcal{P}_l$ of the $\tau$-leaping method may lead to negative solutions \cite{tian2004binomial, cao2005avoiding} which are easily avoided in our method by simply bounding the copy number of reactants as the reaction channels are uncorrelated for reactants in every subset. This treatment is simpler than existing strategies, e.g. the hybrid of $\tau$-leaping method and Gillespie's stochastic simulation algorithm (SSA) \cite{gillespie1977exact} in ref.\cite{cao2005avoiding}, and does not affect the computational efficiency of the $\tau$-leaping method.

\begin{figure}[h]
\centering
\includegraphics[width=1.0\linewidth]{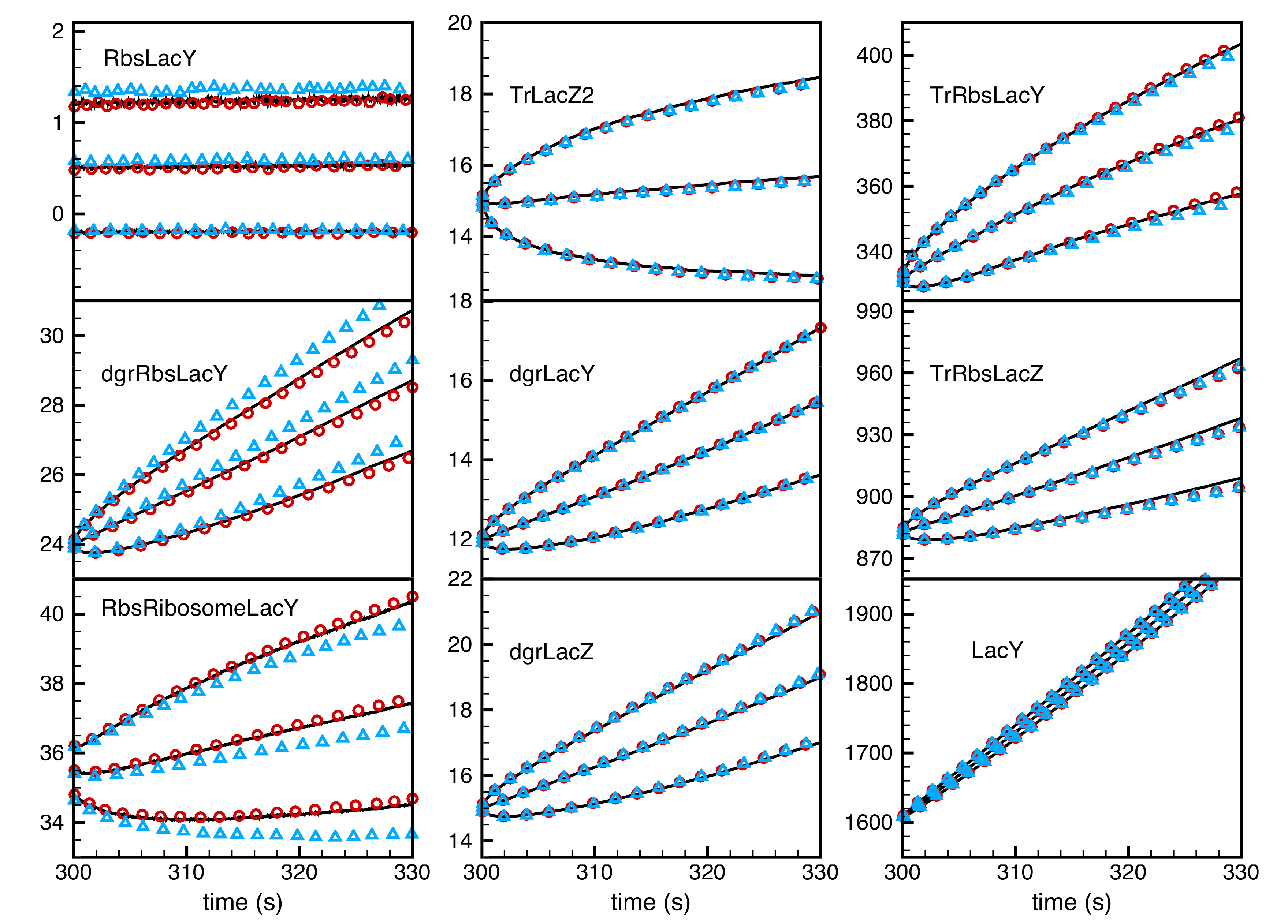}
\caption{Time evolution of means ($\chi$) and standard deviations ($\sigma$) of the species concentration in the LacZ/LacY reaction. The upper, middle, and lower trajectories (lines and symbols) corresponds to the mean ($\chi$) and the bounds $\chi \pm \sigma$, respectively. Numerical results generated by the SSA are considered as the reference solutions and represented by the black solid lines. Starting from the initial condition generated by SSA at $t=300s$, we perform numerical simulations until $t=330s$ by the Lie splitting method (symbols $\circ$) and our network-partition method (symbols $\Delta$). A constant leap time $\tau = 0.00625$ is used and the ensemble average is performed by $10,000$ simulations to obtain the mean and standard deviation values. 
The molecular numbers of RNAP and Ribosome are updated by $\mathcal{N}(35 \mathcal{V}, 3.5^2)$ and $\mathcal{N}(350 \mathcal{V}, 35^2)$, where $\mathcal{V}= 1.0+t/t_g$ is the cell volume and $t_g = 2100s$ is the cell generation time.  Negative solutions are not observed in any of the $10,000$ simulations.}
\label{fig:laczy}
\end{figure}

We consider the LacZ/LacY reaction model \cite{kierzek2002stocks} with $22$ reactions, as shown in Table \hyperref[sec:si]{VII}, to verify the accuracy present method. This case, insipite its small size ($N=22$), show distinct sparse network features and can be which are clustered into $K=5$ subsets. Initially, the population of PLac is $1$ while others are $0$. The numbers of RNAP and Ribosome are updated by $\mathcal{N}(35 \mathcal{V}, 3.5^2)$ and $\mathcal{N}(350 \mathcal{V}, 35^2)$, where $\mathcal{V}= 1.0+t/t_g$ is the cell volume and $t_g = 2100s$ is the cell generation time \cite{kierzek2002stocks}. And the propensities of the reactions $(1,8,9,20)$ in Table \hyperref[sec:si]{VII} are rescaled by $\mathcal{V}$.
First, we simulate this system until $t=300s$ by SSA. Then using this result as the input, $10,000$ simulations are performed in the time interval $[300s,330s]$ by our method and SSA. Mean and standard deviations of molecular numbers are computed. As shown in Fig.\ref{fig:laczy}, the predicted trajectories of our results with $\tau=0.00625s$ agree well with of the SSA results. And no negative molecular number is observed although the number of RbsLacY is very low ($\simeq 1$) where original $\tau$-leaping method easily produces negative solutions \cite{tian2004binomial}.

\subsubsection{Large-scale reactions}
\begin{figure}[h]
%\centering
\hspace{0.5cm} \includegraphics[width=0.6\linewidth]{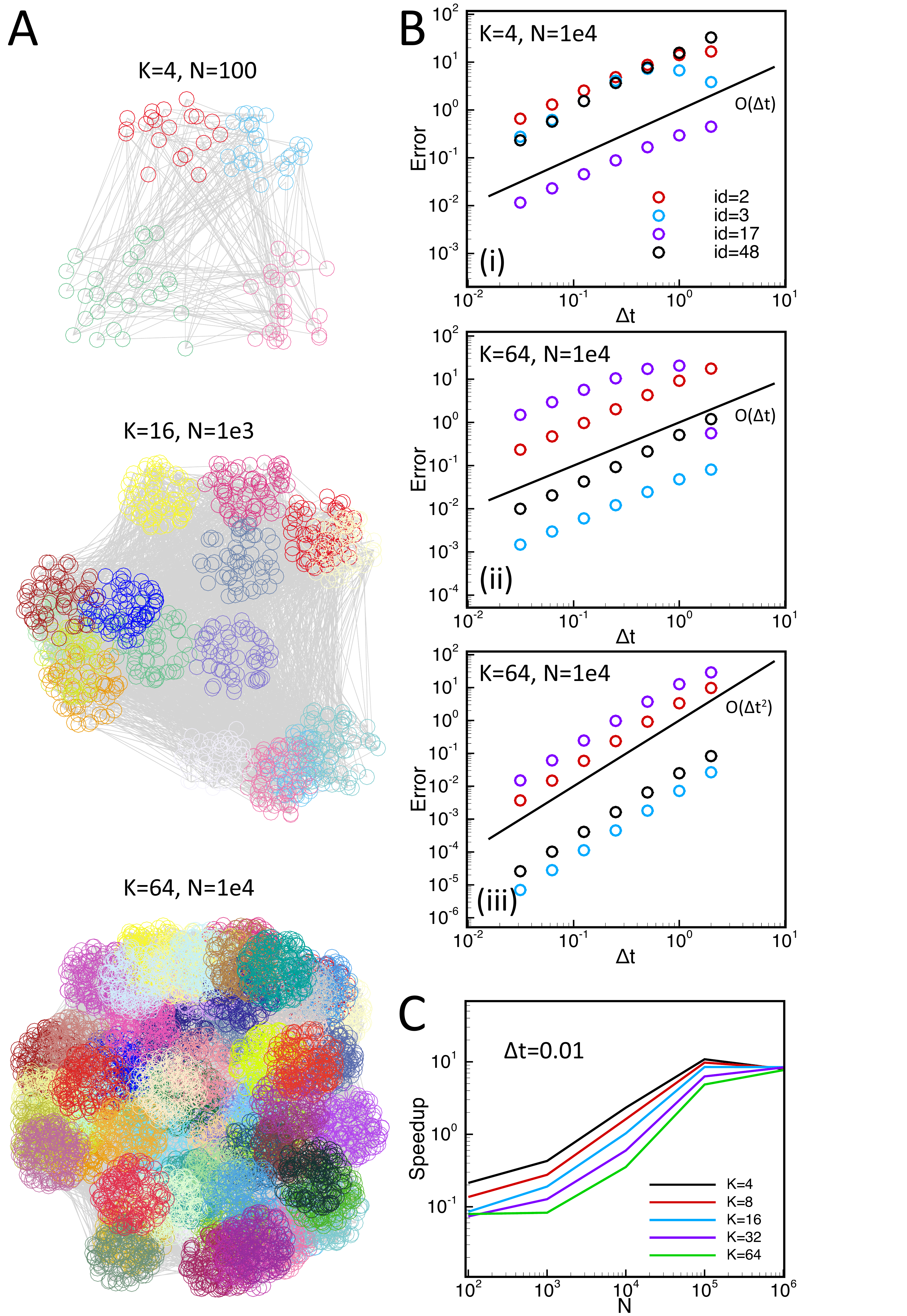}
\caption{Numerical results of large-scale randomly generated reaction systems by the Lie splitting and the network-partition based methods. The elementary reactions are randomly chosen from Table \hyperref[sec:si]{VII} and reaction rate coefficient is $\alpha = \mathcal{U}(10^{-3},1)$. The network structures of reaction systems with $N=100$, $1000$ and $10,000$ and $M=N$ reacting species are shown in (A). The elements belonging to the same subset are represented by the same color, exhibiting clear network structures. The convergence of errors generated by our method are measured at $t=300s$ and plotted in (B) with $K=4$ (i) and $K=64$ (ii and iii), where $K$ is the number of subsets and ``id'' is the index of reacting species. Combining the Strang splitting scheme with our method achieves second-order convergence rate in (B, iii). (C) The speedup of parallel simulations, defined by the ratio of CPU time between the serial simulation and the parallel simulation performed on a 12-core (Intel Xeon CPU E5-2609 v4, 1.70GHz) desktop, approaches the expected value ($12$) when the scale of the reaction is increasing from $N=100$ to $N=10^6$.}
\label{fig:para}
\end{figure}

In this section we extend the application to large-scale reacting systems to test the efficiency of our method. 
The type of each elementary reaction is randomly selected from all types in the LacZ/LacY reaction model and initially we set $\mathbf{X}=\mathbf{0}$. The reaction rate coefficient is given by $\mathcal{U}(10^{-3},1)$ and the deterministic equation (\ref{eq:rre}) are solved by our method with the law of mass action. We increase the size of the reaction system from $N=100$ to $N=10^6$ and set $M=N$, as shown in Fig.\ref{fig:para}A. 

The errors measured at $t=100$ show first-order convergence rate in Fig.\ref{fig:para}B if we use Lie scheme after partition. As expected, a 2nd-order convergence rate is observed if we use Strang scheme. Then we parallize the algorithm and compute the speedup which is the ratio of CPU time for the serial simulation and for the parallel simulations performed on a 12-core desktop. When $N$ is small, the speedup is low as the computational cost of every subset is insufficient. It approaches the expected speedup value ($12$) when the scale is increasing, indicating the parallelization property of our method and thus substantially save the computational time for large-scale computations.

\subsection{Reaction-diffusion on complex networks} \label{sec:rd}
Reaction-diffusion process is the underlying mechanism of many pattern formulations in biological systems. Its behaviour on cellular networks can be used to model early stage morphogenesis \cite{othmer1971instability}. Recently, reaction-diffusion (RD) processes on random networks with size up to $10^3$ nodes show significant difference from the class behaviour \cite{nakao2010turing}. In this section we demonstrate our method can be used to efficiently solve RD processes on large-scale networks.

Consider a RD system defined on a complex network with $N^*$ nodes and $M$ different species. On each node, the local reactions change the concentrations of every species which are diffusively transported to connected nodes. The dynamic behaviour is described by
\begin{eqnarray}\label{eq:rd}
\frac{d \text{X}_{in}}{d t} &=& \sum_{j=1}^N (\nu_{ji} - \mu_{ji}) \text{C}_j(t) + D_i \sum_{m=1}^{N^*} L_{nm} \text{X}_{im}, \nonumber
\end{eqnarray}
where $1 \leq i \leq M$, $1 \leq n \leq N^*$, $D_i$ is the diffusion rate of the species $i$ and $L_{nm} = A_{nm}-d_n \delta_{nm}$ is the network Laplacian matrix.

The Lie operator splitting method can be used for highly dissipative system in particle simulations of fluid mechanics whose diffusion term is handled by decoupling all fluxes for each particle and updating the solution in a pairwise manner \cite{litvinov2010splitting}. Here we consider the flux $\text{F}^i_{nm} = \text{X}_{in}-\text{X}_{im}$ as the element and cluster it by the same way for chemical reactions, i.e., any two fluxes of each subset do not flow into or out of the same node. For the species $i$ and the subset $S_k$, the concentrations of node $n$ and $m$ are updated by $\text{X}_{in} = \text{X}_{in} - D_i\, \Delta t\, \text{F}^i_{nm}$ and $\text{X}_{im} = \text{X}_{im} - D_i\, \Delta t\, \text{F}^i_{nm}$, respectively. Thus positivity of $\mathbf{X}$ can be ensured by limiting the flux $\text{F}^i_{nm} = \max\left( -\text{F}_{im}/(D_i\, \Delta t), \min \left( \text{F}^i_{nm}, \text{F}_{in}/(D_i\, \Delta t) \right) \right) $.

\begin{figure}[h]
\centering
\includegraphics[width=0.8\linewidth]{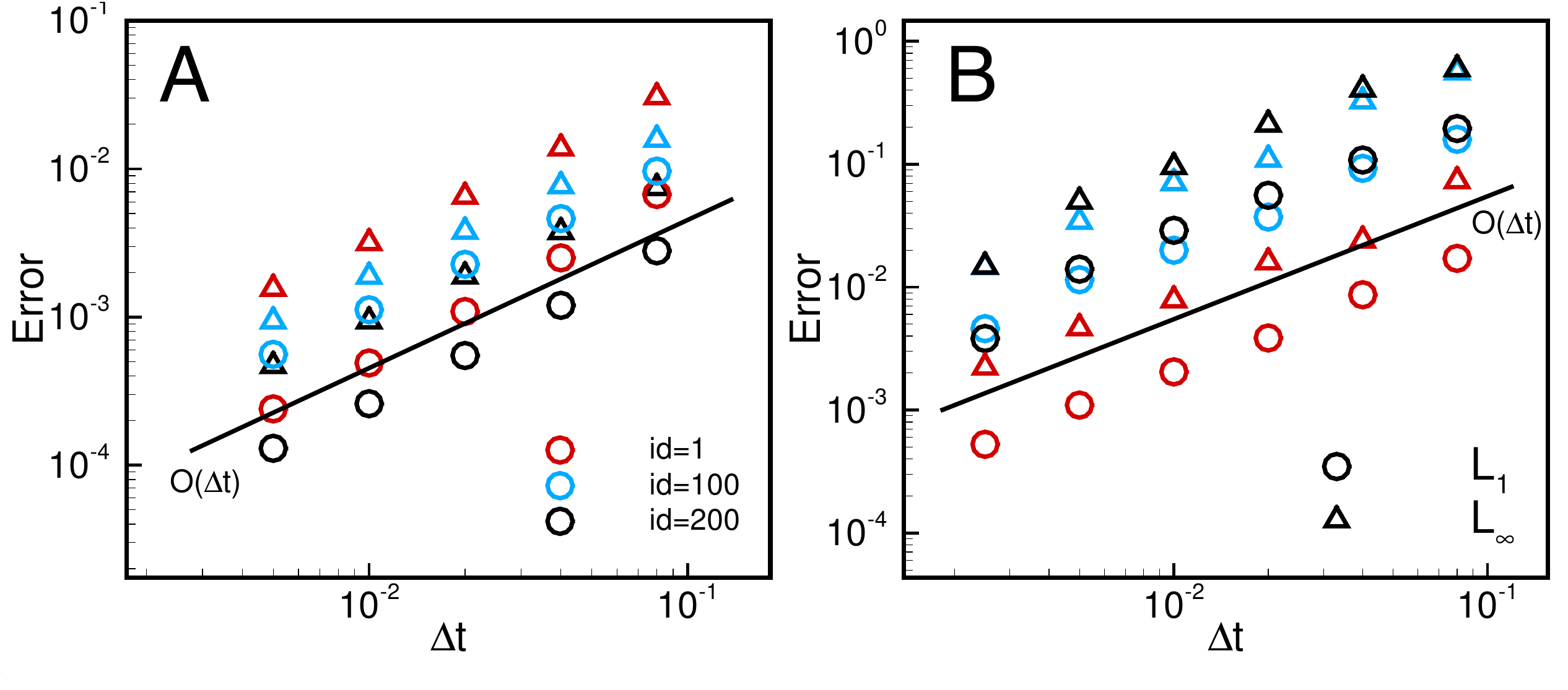}
\caption{Numerical results of reaction-diffusion processes on a small scale-free network with $N^*=200$ nodes and a mean degree $\bar{d}=12$. The error convergence results of pure diffusion are sketched in (A). The nodes are sorted in increasing order of their degree, i.e., larger index value (``id'') indicates larger node degree. The results of our method (symbols $\circ$) are compared to those of the Lie splitting method (symbols $\Delta$). The convergence of L1 and L$_{\infty}$ errors of a 3-speceis reaction-diffusion are sketched in (B).}
\label{fig:rdsmall}
\end{figure}

We first test pure diffusion process on a small scale-free network \cite{barabasi1999emergence} with $N^*=200$ nodes and a mean degree $\bar{d}=12$. The diffusion rate $D=2.04$ and initially sate is $\text{X} = 10 + 10 \mathcal{U}(-0.01,0.01)$. As shown in Fig.\ref{fig:rdsmall}A, errors are measured when the system achieves the equilibrium state and exhibit first-order convergence. And our method shows better accuracy than the Lie splitting method.
If we increase $D$, the numerical instabiles are observed if we use the 2nd-order Runge-Kutta scheme with large time-steps. This issue is addressed by our method, indicating that our method, like that in ref.\cite{litvinov2010splitting}, can handle highly dissipative system efficiently.
Then the Brusselator reaction model \cite{hata2014dispersal},
\begin{eqnarray}\label{eq:bru}
&&\frac{d \text{X}_{1n}}{d t} = 1-3.9 \text{X}_{1n}+(\text{X}_{1n})^2 \text{X}_{2n}-\text{X}_{1n}+ \text{X}_{3n}, \nonumber \\
\nonumber
&&\frac{d \text{X}_{2n}}{d t} = 2.9 \text{X}_{1n}-(\text{X}_{1n})^2 \text{X}_{2n}, \quad \frac{d \text{X}_{3n}}{d t} = \text{X}_{1n}-\text{X}_{3n}
\end{eqnarray}
is added for every node. The corresponding steady state is $\bar{\mathbf{X}}= \left\lbrace 1,2.9,1 \right\rbrace$ and will be destabilized under specific diffusion rate, leading to the Turing instability. A small perturbation $0.01\bar{\mathbf{X}}$ is imposed on $\bar{\mathbf{X}}$ initially and the diffusion rates are set as $D_1=D_2=7\times 10^{-3}$ and $D_3=0.161$. The convergence results in Fig.\ref{fig:rdsmall}B exhibit first order rate for $L_1$ and $L_{\infty}$ norms.

\begin{figure}[h]
\centering
\includegraphics[width=1.0\linewidth]{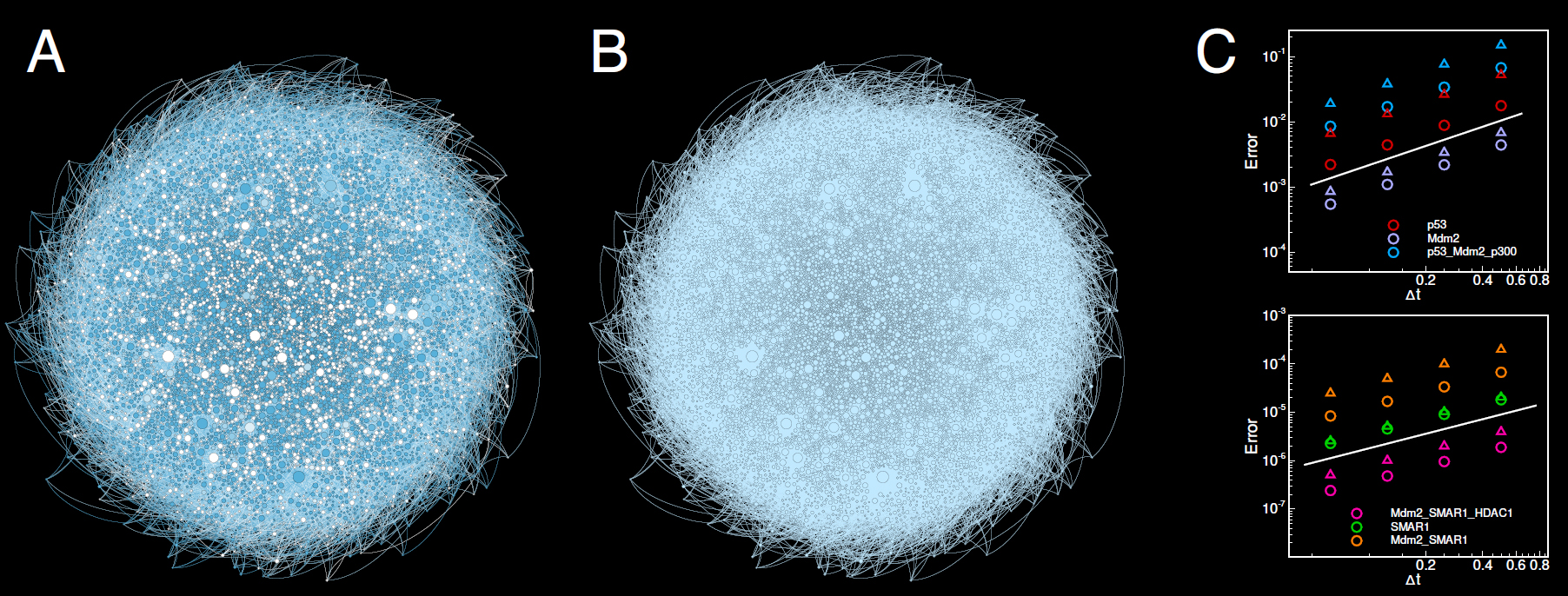}
\caption{Numerical results of large-scale RD processes. Spatial distribution of p53 for low diffusion rate and high diffusion rate are shown in (A) and (B), respectively. The convergence results ($\circ$) are shown in (C) and compared to the results ({\tiny $\triangle$}) of Lie operator splitting.}
\label{fig:reacdiffu}
\end{figure}

Then we study a large-scale RD system defined on a scale-free network with $10^4$ nodes with $\bar{d}=10$. The p53-SMAR1 regulatory reaction model is considered for the reaction term. 
The initial condition of $\mathbf{X}$ and the diffusion rate are listed in Table \hyperref[sec:si]{X}.
The $49986$ fluxes of this system are clustered into $569$ subsets and the partition of the chemical reaction is $S^{(0)}$ in Table \hyperref[sec:si]{IV}. The initial condition of $\mathbf{X}$ and diffusion rates are listed in Table . 
When the diffusion rate is low ($0.1\%$ of the rates in Table), the distribution of p53 is highly heterogeneous (Fig.\ref{fig:reacdiffu}A) and will be homogenized by large diffusion rate (Fig.\ref{fig:reacdiffu}B). The convergence analysis is shown in Fig.\ref{fig:reacdiffu}C which indicates our method is 1st-order accurate in this case and more accurate than Lie splitting method.

\section{Conclusion}\label{sec:con}
We present a numerical method based on network partition and operator splitting to solve large-scale complex processes. It overcomes numerical difficulties of conventional methods for large-scale problems, including convergence issue, memory overload, conservation, robustness and efficiency. Our method tries to exploit the network structure of large-scale system and utilize the sparsity of it to partition the system into smaller subproblems which are easy for numerical computations. In this way, large computational costs due to stiffness, high dissipation, and large number of involved species for chemical-reaction and reaction-diffusion processes are significantly reduced. This method is easy to be parallelized and shows good convergence properties. 
A range of applications demonstrate the flexibility, modularity, robustness, and versatility of our methods, indicating that it is suitable for solving complex problems involving a large number of elementary processes.

\section*{Materials and Methods}
The source code of the numerical method in this paper has been uploaded to \href{https://gitlab.com}{https://gitlab.com} and the mechanism is detailed in \hyperref[sec:si]{Supporting Information}.
\section*{Acknowledgment}
This research is supported by China Scholarship Council (No. 201306290030), National Natural Science Foundation of China (No. 11628206), and Deutsche
Forschungsgemeinschaft (HU 1527/6-1).

\bibliographystyle{abbrv}
\bibliography{pnas-sample}% Produces the bibliography via BibTeX.

\begin{thebibliography}{10}

\bibitem{asllani2014theory}
M.~Asllani, J.~D. Challenger, F.~S. Pavone, L.~Sacconi, and D.~Fanelli.
\newblock The theory of pattern formation on directed networks.
\newblock {\em Nat Commun}, 5:4517, 2014.

\bibitem{balcan2009multiscale}
D.~Balcan, V.~Colizza, B.~Gon{\c{c}}alves, H.~Hu, J.~J. Ramasco, and
  A.~Vespignani.
\newblock Multiscale mobility networks and the spatial spreading of infectious
  diseases.
\newblock {\em Proc Natl Acad Sci USA}, 106(51):21484--21489, 2009.

\bibitem{barabasi1999emergence}
A.-L. Barab{\'a}si and R.~Albert.
\newblock Emergence of scaling in random networks.
\newblock {\em science}, 286(5439):509--512, 1999.

\bibitem{brown1989vode}
P.~N. Brown, G.~D. Byrne, and A.~C. Hindmarsh.
\newblock Vode: A variable-coefficient {ODE} solver.
\newblock {\em SIAM J Sci Comput}, 10(5):1038--1051, 1989.

\bibitem{cao2005avoiding}
Y.~Cao, D.~T. Gillespie, and L.~R. Petzold.
\newblock Avoiding negative populations in explicit {P}oisson tau-leaping.
\newblock {\em J Chem Phys}, 123(5):054104, 2005.

\bibitem{chorin1997numerical}
A.~J. Chorin.
\newblock A numerical method for solving incompressible viscous flow problems.
\newblock {\em J Comput Phys}, 135(2):118--125, 1997.

\bibitem{colizza2007reaction}
V.~Colizza, R.~Pastor-Satorras, and A.~Vespignani.
\newblock Reaction--diffusion processes and metapopulation models in
  heterogeneous networks.
\newblock {\em Nat Phys}, 3(4):276--282, 2007.

\bibitem{descombes2001convergence}
S.~Descombes.
\newblock Convergence of a splitting method of high order for
  reaction-diffusion systems.
\newblock {\em Math Comput}, 70(236):1481--1501, 2001.

\bibitem{deuflhard1986efficient}
P.~Deuflhard and U.~Nowak.
\newblock Efficient numerical simulation and identification of large chemical
  reaction systems.
\newblock {\em Ber Bunsenges Phys Chem}, 90(11):940--946, 1986.

\bibitem{diegelmann2017three}
F.~Diegelmann, S.~Hickel, and N.~A. Adams.
\newblock Three-dimensional reacting shock--bubble interaction.
\newblock {\em Combust Flame}, 181:300--314, 2017.

\bibitem{gillespie1977exact}
D.~T. Gillespie.
\newblock Exact stochastic simulation of coupled chemical reactions.
\newblock {\em J Phys Chem}, 81(25):2340--2361, 1977.

\bibitem{gillespie2001approximate}
D.~T. Gillespie.
\newblock Approximate accelerated stochastic simulation of chemically reacting
  systems.
\newblock {\em J Chem Phys}, 115(4):1716--1733, 2001.

\bibitem{glowinski2017splitting}
R.~Glowinski, S.~J. Osher, and W.~Yin.
\newblock {\em Splitting Methods in Communication, Imaging, Science, and
  Engineering}.
\newblock Springer, 2017.

\bibitem{goldstein2009split}
T.~Goldstein and S.~Osher.
\newblock The split bregman method for {L}1-regularized problems.
\newblock {\em SIAM J Imaging Sci}, 2(2):323--343, 2009.

\bibitem{gou2010dynamic}
X.~Gou, W.~Sun, Z.~Chen, and Y.~Ju.
\newblock A dynamic multi-timescale method for combustion modeling with
  detailed and reduced chemical kinetic mechanisms.
\newblock {\em Combust Flame}, 157(6):1111--1121, 2010.

\bibitem{hata2014dispersal}
S.~Hata, H.~Nakao, and A.~S. Mikhailov.
\newblock Dispersal-induced destabilization of metapopulations and oscillatory
  {T}uring patterns in ecological networks.
\newblock {\em Sci Rep}, 4:3585, 2014.

\bibitem{higham2008modeling}
D.~J. Higham.
\newblock Modeling and simulating chemical reactions.
\newblock {\em SIAM Rev}, 50(2):347--368, 2008.

\bibitem{karlebach2008modelling}
G.~Karlebach and R.~Shamir.
\newblock Modelling and analysis of gene regulatory networks.
\newblock {\em Nat Rev Mol Cell Biol}, 9(10):770--780, 2008.

\bibitem{kierzek2002stocks}
A.~M. Kierzek.
\newblock {STOCKS}: {STOChastic Kinetic Simulations} of biochemical systems
  with {G}illespie algorithm.
\newblock {\em Bioinformatics}, 18(3):470--481, 2002.

\bibitem{lafon2006diffusion}
S.~Lafon and A.~B. Lee.
\newblock Diffusion maps and coarse-graining: {A} unified framework for
  dimensionality reduction, graph partitioning, and data set parameterization.
\newblock {\em IEEE Trans Pattern Anal Mach Intell}, 28(9):1393--1403, 2006.

\bibitem{leveque1990study}
R.~J. LeVeque and H.~C. Yee.
\newblock A study of numerical methods for hyperbolic conservation laws with
  stiff source terms.
\newblock {\em J Comput Phys}, 86(1):187--210, 1990.

\bibitem{lie1888theorie}
S.~Lie and F.~Engel.
\newblock Theorie der {T}ransformationsgruppen.
\newblock 1888.

\bibitem{litvinov2010splitting}
S.~Litvinov, M.~Ellero, X.~Hu, and N.~Adams.
\newblock A splitting scheme for highly dissipative smoothed particle dynamics.
\newblock {\em J Comput Phys}, 229(15):5457--5464, 2010.

\bibitem{lu2005directed}
T.~Lu and C.~K. Law.
\newblock A directed relation graph method for mechanism reduction.
\newblock {\em Proc Combust Inst}, 30(1):1333--1341, 2005.

\bibitem{lu2009toward}
T.~Lu and C.~K. Law.
\newblock Toward accommodating realistic fuel chemistry in large-scale
  computations.
\newblock {\em Prog Energy Combust Sci}, 35(2):192--215, 2009.

\bibitem{malik2017control}
M.~Z. Malik, M.~J. Alam, R.~Ishrat, S.~M. Agarwal, and R.~B. Singh.
\newblock Control of apoptosis by {SMAR} 1.
\newblock {\em Mol Biosyst}, 13(2):350--362, 2017.

\bibitem{nakao2010turing}
H.~Nakao and A.~S. Mikhailov.
\newblock Turing patterns in network-organized activator-inhibitor systems.
\newblock {\em Nat Phys}, 6(7):544--550, 2010.

\bibitem{newman2012communities}
M.~E. Newman.
\newblock Communities, modules and large-scale structure in networks.
\newblock {\em Nat Phys}, 8(1):25, 2012.

\bibitem{nguyen2009mass}
K.~Nguyen, A.~Caboussat, and D.~Dabdub.
\newblock Mass conservative, positive definite integrator for atmospheric
  chemical dynamics.
\newblock {\em Atmos Environ}, 43(40):6287--6295, 2009.

\bibitem{oran2005numerical}
E.~S. Oran and J.~P. Boris.
\newblock {\em Numerical simulation of reactive flow}.
\newblock Cambridge University Press, 2005.

\bibitem{othmer1971instability}
H.~G. Othmer and L.~Scriven.
\newblock Instability and dynamic pattern in cellular networks.
\newblock {\em J Theor Biol}, 32(3):507--537, 1971.

\bibitem{rain2001protein}
J.-C. Rain, L.~Selig, H.~De~Reuse, V.~Battaglia, C.~Reverdy, S.~Simon,
  G.~Lenzen, F.~Petel, J.~Wojcik, V.~Sch{\"a}chter, et~al.
\newblock The protein--protein interaction map of helicobacter pylori.
\newblock {\em Nature}, 409(6817):211--215, 2001.

\bibitem{schwer2003adaptive}
D.~A. Schwer, P.~Lu, and W.~H. Green.
\newblock An adaptive chemistry approach to modeling complex kinetics in
  reacting flows.
\newblock {\em Combust Flame}, 133(4):451--465, 2003.

\bibitem{slack1977investigation}
M.~Slack and A.~Grillo.
\newblock Investigation of hydrogen-air ignition sensitized by nitric oxide and
  by nitrogen dioxide.
\newblock 1977.

\bibitem{speth2013balanced}
R.~L. Speth, W.~H. Green, S.~MacNamara, and G.~Strang.
\newblock Balanced splitting and rebalanced splitting.
\newblock {\em SIAM J Numer Anal}, 51(6):3084--3105, 2013.

\bibitem{strang1968construction}
G.~Strang.
\newblock On the construction and comparison of difference schemes.
\newblock {\em SIAM J Numer Anal}, 5(3):506--517, 1968.

\bibitem{thalhammer2008high}
M.~Thalhammer.
\newblock High-order exponential operator splitting methods for time-dependent
  {S}chr{\"o}dinger equations.
\newblock {\em SIAM J Numer Anal}, 46(4):2022--2038, 2008.

\bibitem{tian2004binomial}
T.~Tian and K.~Burrage.
\newblock Binomial leap methods for simulating stochastic chemical kinetics.
\newblock {\em J Chem Phys}, 121(21):10356--10364, 2004.

\bibitem{wagstaff2001constrained}
K.~Wagstaff, C.~Cardie, S.~Rogers, S.~Schr{\"o}dl, et~al.
\newblock Constrained k-means clustering with background knowledge.
\newblock In {\em Proceedings of the Eighteenth International Conference on
  Machine Learning}, volume~1, pages 577--584, 2001.

\end{thebibliography}

\clearpage

\section*{Supporting Information (SI)} \label{sec:si}
\subsection{The mechanism of chemical reactions and the partition results}
We provide the reaction mechanism for the cases used in the main text. First the p53-SMAR1 regulatory model that has $35$ elementary reactions is listed in Table \ref{table:p53_model}. Its network is partitioned by $S^{(0)}$, $S^{(p)}$ and $S^{*,(0)}$, as shown in Table \ref{table:p53_partition}. The hydrogen-oxygen reaction with 9 species and 23 reversible reactions is detailed in Table \ref{table:h2_model} and the corresponding partition result is shown in Table \ref{table:partition_ho}. The LacZ/LacY reaction model contains 22 reactions and 23 species, as shown in Table \ref{table:LacZY_model}. The SSA results at $t=300s$ are used for initial condition in Table \ref{table:lacZY_ini}. The 22 elementary reactions are partitioned into $5$ subsets in Table \ref{table:partition_LacZY}. 
\subsection{Numerical details}
The thermochemical properties of the involved species can be obtained by empirical equations for fitting thermodynamic functions as
\begin{align}\label{eq:entro}
&\frac{c_p}{\text{R}} = \sum_{n=-2}^4 a_{n+3} \text{T}^n, \nonumber \\
&\frac{h}{\text{RT}}  = - \frac{a_1}{\text{T}^2}   + a_2 \frac{\text{ln}\text{T}}{\text{T}} + 
\sum_{n=0}^4 \frac{a_{n+3} \text{T}^n}{n+1} + \frac{b_1}{\text{T}}, \nonumber \\
&\frac{s}{\text{R}}   = -\frac{a_1}{2\text{T}^2}  - \frac{2}{\text{T}}  + a_3 \text{ln}\text{T} + 
\sum_{n=1}^4 \frac{a_{n+3} \text{T}^n}{n} + b_2, \nonumber
\end{align}
over a wide temperature range, where coefficients $a_1$ to $a_7$ and $b_{1,2}$ are tabulated in Table \ref{thermodata}. For each species, the first row applies for a temperature range from 1000K to 6000 K and second row is used for temperature below 1000K.  
The temperature $\text{T}$ of the mixture is updated by iteratively solving the thermodynamics relation 
\begin{align}
h = e + \frac{p}{\rho} = e + \sum_{m=1}^M y_m \text{R}_m \rho \text{T},   \nonumber
\end{align}
considering the internal energy $e$ is constant during the adiabatic process of $\Delta t$.

\begin{table}%[tbhp]
\centering
\caption{Reaction mechanism for p53-SMAR1 regulatory biochemical network (part I).}
\begin{tabular}{llc}
Id			& elementary reaction  & reaction rate coefficient \\
\hline
1			& p53 + Mdm2$\_$p300 $\rightarrow$ $\emptyset$ & 8.25e-4 \\
2			& Mdm2$\_$mRNA $\rightarrow$ Mdm2$\_$mRNA + Mdm2 & 4.95e-4 \\
3			& p53 $\rightarrow$ p53 + Mdm2$\_$mRNA & 1e-4 \\
4			& Mdm2$\_$mRNA $\rightarrow$ $\emptyset$ & 1e-4 \\
5			& Mdm2 $\rightarrow$ $\emptyset$ & 4.33e-4 \\
6			& $\emptyset$ $\rightarrow$ p53 & 0.078 \\
7			& p53$\_$Mdm2 $\rightarrow$ Mdm2 & 8.25e-4 \\
8			& p53 + Mdm2 $\rightarrow$ p53$\_$Mdm2 & 11.55e-4 \\
9			& p53$\_$Mdm2 $\rightarrow$ p53 + Mdm2 & 11.55e-6 \\
10			& ATM$\_$I $\rightarrow$ ATM$\_$A & 1e-4 \\
11			& ATM$\_$A $\rightarrow$ ATM$\_$I & 5e-4 \\
12			& p53 + ATM$\_$A $\rightarrow$ ATM$\_$A + p53$_p$ & 5e-4 \\
13			& p53$_p$ $\rightarrow$ p53 & 0.5 \\
14			& p300 $\rightarrow$ $\emptyset$ & 1e-4 \\
15			& p53$_p$ + p300 $\rightarrow$ p53$_p\_$p300 & 1e-4 \\
16			& p53$_p\_$p300 $\rightarrow$ p53$\_$A & 1e-4 \\
17			& p53$\_$A + Mdm2$\_$SMAR1$\_$HDAC1 $\rightarrow$ p53 & 1e-5 \\
\hline
\end{tabular}
\label{table:p53_model}
\end{table}
\begin{table}%[tbhp]
\centering
\caption{Reaction mechanism for p53-SMAR1 regulatory biochemical network (part II).}
\begin{tabular}{llc}
Id			& elementary reaction  & reaction rate coefficient \\
\hline
17			& p53$\_$A + Mdm2$\_$SMAR1$\_$HDAC1 $\rightarrow$ p53 & 1e-5 \\
18			& Mdm2 + SMAR1 $\rightarrow$ Mdm2$\_$SMAR1 & 2e-4 \\
19			& p53$\_$Mdm2 + p300 $\rightarrow$ p53$\_$Mdm2$\_$p300 & 5e-4 \\
20			& Mdm2 + p300 $\rightarrow$  Mdm2$\_$p300 & 5e-4 \\
21			& p53$\_$Mdm2$\_$p300 $\rightarrow$  Mdm2 + p53$_p\_$p300 & 1e-4 \\
22			& HDAC1$\rightarrow$  $\emptyset$ & 1e-4 \\
23			& $\emptyset$ $\rightarrow$ p300  & 0.1 \\
24			& $\emptyset$ $\rightarrow$ HDAC1  & 2e-4 \\
25			& HDAC1 +  Mdm2$\_$SMAR1 $\rightarrow$ Mdm2$\_$SMAR1$\_$HDAC1  & 1e-4 \\
26			& $\emptyset$ $\rightarrow$ SMAR1  & 0.08 \\
27			& SMAR1 $\rightarrow$  $\emptyset$  & 1e-4 \\
28			& Mdm2$\_$SMAR1 $\rightarrow$  $\emptyset$  & 2e-4 \\
29			& p53 + SMAR1 $\rightarrow$  p53$_p$  & 1e-4 \\
30			& p53$\_$Mdm2 + SMAR1 $\rightarrow$  p53$\_$Mdm2$\_$SMAR1  & 1e-3 \\
31			& p53$\_$Mdm2$\_$SMAR1 $\rightarrow$  p53$_p$ + Mdm2$\_$SMAR1 & 1e-3 \\
32			& Mdm2 + HDAC1 $\rightarrow$  Mdm2$\_$HDAC1 & 2e-3 \\
33			& Mdm2$\_$HDAC1 + p53$\_$A $\rightarrow$ p53 & 5 \\
34			& p53$_p\_$p300 + SMAR1 $\rightarrow$ p53 + SMAR1 & 1e-4 \\
35			& p300 + SMAR1 $\rightarrow$ SMAR1 & 0.5 \\
\hline
\end{tabular}
\end{table}
\begin{table}%[tbhp]
\centering
\caption{Partition for the p53-SMAR1 regulatory biochemical network.}
\label{table:p53_partition}
\begin{tabular}{llll}
subset			& elements of $S^{(0)}$ & elements of $S^{(p)}$ & elements of $S^{*,(0)}$\\
\hline
$S_1$			& \{1, 2, 10, 14, 16, 22, 26, 28\} & \{1, 24, 35\} & \{1, 2, 10, 14, 16, 22, 26, 28\} \\
$S_2$			& \{3, 5, 11, 15, 24, 27\}  & \{3, 5, 16\} &  \{3, 5, 11, 15, 24, 27\}\\
$S_3$			& \{4, 6, 7, 23\} & \{6, 31, 32\} & \{4, 6, 7, 23, 25\}\\
$S_4$			& \{8, 31, 35\} & \{4, 8, 22\} & \{8, 31, 35\}\\
$S_5$			& \{9\} & \{9, 15, 27, 28\} & \{9, 17, 21, 34\}\\
$S_6$			& \{12, 18, 19\} & \{12, 20, 30\} & \{12, 18, 19\}\\
$S_7$			& \{13, 20, 30\} & \{7, 10, 13, 26\} & \{13, 20, 30\}\\
$S_8$			& \{17, 21\} & \{17, 18, 23\} & \{29, 32\} \\
$S_9$			& \{29, 32\} & \{19, 25, 29\} & \{33\}\\
$S_{10}$			& \{33\} & \{11, 21, 33\} & $\emptyset$\\
$S_{11}$			& \{34\} & \{2, 14, 34\} & $\emptyset$\\
\hline
\end{tabular}
\end{table}

\begin{table}
\centering
\caption{H2-air mixture reaction mechanism for combustion.}
\begin{tabular}{lllll}
Id		& elementary reaction & $A$ & $B$ & $E_a$  \\
\hline
1,2		& $\text{H} + \text{O}_2 \Longleftrightarrow \text{OH} + \text{O}$ & 1.91e+14 & 0.0 & 16.44 \\
3,4		& $\text{H}_2 + \text{O} \Longleftrightarrow \text{H} + \text{OH}_2 $ & 5.08e+04 & 2.67 & 6.292 \\ 
5,6		& $\text{H}_2 + \text{OH} \Longleftrightarrow \text{H} + \text{H}_2\text{O} $ & 2.16e+08 & 1.51 & 3.43 \\
7,8		& $\text{O} + \text{H}_2\text{O} \Longleftrightarrow \text{OH} + \text{OH} $ & 2.97e+06 & 2.02 & 13.4 \\ 
9,10*	& $\text{H}_2 + \text{M} \Longleftrightarrow \text{H} + \text{H} + \text{M} $ & 4.57e+19 & -1.4 & 105.1 \\ 
11,12*	& $\text{O} + \text{O} + \text{M} \Longleftrightarrow \text{O}_2 + \text{M} $ & 6.17e+15 & -0.5 & 0.0 \\
13,14*	& $\text{H} + \text{O} + \text{M} \Longleftrightarrow \text{OH} + \text{M} $ & 4.72e+18 & -1.0 & 0.0 \\
15,16**	& $\text{H} + \text{OH} + \text{M} \Longleftrightarrow \text{H}_2\text{O} + \text{M} $ & 4.50e+22 & -2.0 & 0.0 \\
17,18***& $\text{H} + \text{O}_2 + \text{M} \Longleftrightarrow \text{H}\text{O}_2 + \text{M} $ & 3.48e+16 & -0.41 & -1.12\\
19,20***& $\text{H} + \text{O}_2 \Longleftrightarrow \text{H}\text{O}_2 $ & 1.48e+12 & 0.60 & 0.0 \\
21,22	& $\text{H} + \text{HO}_2 \Longleftrightarrow \text{H}_2 + \text{O}_2 $ & 1.66e+13 & 0.0 & 0.82 \\
23,24	& $\text{H} + \text{HO}_2 \Longleftrightarrow \text{OH} + \text{OH} $ & 7.08e+13 & 0.0 & 0.3 \\
25,26	& $\text{HO}_2 + \text{O} \Longleftrightarrow \text{OH} + \text{O}_2 $ & 3.25e+13 & 0.0 & 0.0 \\
27,28	& $\text{OH} + \text{HO}_2 \Longleftrightarrow \text{H}_2\text{O} + \text{O}_2 $ & 2.89e+13 & 0.0 & -0.5 \\
29,30	& $\text{H}\text{O}_2 + \text{H}\text{O}_2 \Longleftrightarrow \text{H}_2\text{O}_2 + \text{O}_2$ & 4.20e+14 & 0.0 & 11.98 \\
31,32	& $\text{H}\text{O}_2 + \text{H}\text{O}_2 \Longleftrightarrow \text{H}_2\text{O}_2 + \text{O}_2$ & 1.30e+11 & 0.0 & -1.629 \\
33,34*	& $\text{H}_2\text{O}_2 + \text{M} \Longleftrightarrow \text{OH} + \text{OH} + \text{M} $ & 1.27e+17 & 0.0 & 45.5 \\
35,36*	& $\text{H}_2\text{O}_2 \Longleftrightarrow \text{OH} + \text{OH} $ & 2.95e+14 & 0.0 & 48.4 \\
37,38	& $\text{H}_2\text{O}_2 + \text{H} \Longleftrightarrow \text{H}_2\text{O} + \text{OH} $ & 2.41e+13 & 0.0 & 3.97 \\
39,40	& $\text{H}_2\text{O}_2 + \text{H} \Longleftrightarrow \text{H}_2 + \text{HO}_2 $ & 6.03e+13 & 0.0 & 7.95 \\
41,42	& $\text{H}_2\text{O}_2 + \text{O} \Longleftrightarrow \text{OH} + \text{HO}_2 $ & 9.55e+06 & 2.0 & 3.97 \\
43,44	& $\text{H}_2\text{O}_2 + \text{OH} \Longleftrightarrow \text{H}_2\text{O} + \text{HO}_2 $ & 1.00e+12 & 0.0 & 0.0 \\
45,46	& $\text{H}_2\text{O}_2 + \text{OH} \Longleftrightarrow \text{H}_2\text{O} + \text{HO}_2 $ & 5.80e+14 & 0.0 & 9.56 \\
\hline

\multicolumn{5}{l}{%
  \begin{minipage}{10cm}
  Third-body collision coefficiencies (default value is 1.0) in reactions with M: *  $\text{H}_2\text{O}=12.0$, $\text{H}_2=2.5$; ** $\text{H}_2\text{O}=12.0$, $\text{H}_2=0.73$; ***$\text{H}_2\text{O}=14.0$, $\text{H}_2=1.3$.
  \end{minipage}%
}\\
\end{tabular}
\label{table:h2_model}
\end{table}

\begin{table}%[tbhp]
\centering
\caption{Partition for the hydrogen-oxygen reaction.}
\label{table:partition_ho}
\begin{tabular}{llllllll}
	 			& $S_1$ & $S_2$ & $S_3$ & $S_4$ & $S_5$ & $S_6$ & $S_7$ \\ 	
\hline
elements			& \{1\}  & \{2\} & \{3, 29\} & \{4,30\} & \{5, 11\} & \{6, 12\} & \{7, 9, 31\} \\
	 			& $S_8$ & $S_9$ & $S_{10}$ & $S_{11}$ & $S_{12}$ & $S_{13}$  & $S_{14}$ \\
\hline
elements			&  \{8, 32\} & \{13\} & \{14\} & \{15\} & \{16\} & \{17, 33\} & \{18, 34\}  \\	
				& $S_{15}$    & $S_{16}$    & $S_{17}$ & $S_{18}$ & $S_{19}$ & $S_{20}$& $S_{21}$\\
\hline
elements			& \{19, 35\}  & \{20, 36\}  & \{21\}   & \{22\}   & \{23\}   & \{24\}  & \{25\}\\
				& $S_{22}$    & $S_{23}$    & $S_{24}$ & $S_{25}$ & $S_{26}$ & $S_{27}$& $S_{28}$\\
\hline
elements			& \{26\}  & \{27\}  & \{28\}   & \{37\}   & \{38\}   & \{39\}  & \{40\}\\
				& $S_{29}$    & $S_{30}$    & $S_{31}$ & $S_{32}$ & $S_{33}$ & $S_{34}$& \\
\hline
elements			& \{41\}  & \{42\}  & \{43\}   & \{44\}   & \{45\}   & \{46\}  & \\
\hline
\end{tabular}
\end{table}
\begin{table*}%[tbhp]
\centering
\caption{Reaction mechanism for the LacZ/LacY reaction.}
\begin{tabular}{llc}
id			& elementary reaction  & reaction rate coefficient \\
\hline
1			& PLac + RNAP $\rightarrow$ PLacRNAP & 0.17 \\
2			& PLacRNAP $\rightarrow$ PLac + RNAP & 10 \\
3			& PLacRNAP $\rightarrow$ TrLacZ1 & 1 \\
4			& TrLacZ1 $\rightarrow$ RbsLacZ + PLac + TrLacZ2 & 1 \\
5			& TrLacZ2 $\rightarrow$ TrLacY2 & 0.015 \\
6			& TrLacY1 $\rightarrow$ RbsLacY + TrLacY2 & 1 \\
7			& TrLacY2 $\rightarrow$ RNAP & 0.36 \\
8			& Ribosome + RbsLacZ $\rightarrow$ RbsribosomeLacZ & 0.17 \\
9			& Ribosome + RbsLacY $\rightarrow$ RbsribosomeLacY & 0.17 \\
10			& RbsribosomeLacZ $\rightarrow$ Ribosome + RbsLacZ & 0.45 \\
11			& RbsribosomeLacY $\rightarrow$ Ribosome + RbsLacY & 0.45 \\
12			& RbsribosomeLacZ $\rightarrow$ TrRbsLacZ + RbsLacZ & 0.4 \\
13			& RbsribsomeLacY $\rightarrow$ TrRbsLacY + RbsLacY & 0.4 \\
14			& TrRbsLacZ $\rightarrow$ LacZ & 0.015 \\
15			& TrRbsLacZ $\rightarrow$ LacY & 0.036 \\
16			& LacZ $\rightarrow$ dgrLacZ & 6.42e-5 \\
17			& LacY $\rightarrow$ dgrLacY & 6.42e-5 \\
18			& RbsLacZ $\rightarrow$ dgrLacY & 0.3 \\
19			& RbsLacZ $\rightarrow$ dgrRbsLacY & 0.3 \\
20			& LacZ + lactose $\rightarrow$ LacZlactose & 9.52e-5 \\
21			& LacZlactose $\rightarrow$ product + LacZ & 431 \\
22			& LacY $\rightarrow$ lactose + LacY & 14 \\
\hline
\end{tabular}
\label{table:LacZY_model}
\end{table*}
\begin{table*}%[tbhp]
\centering
\caption{Initial population of reacting species of the LacZ/LacY reaction.} \label{table:lacZY_ini}
\begin{tabular}{llc}
		id			& reacting species  & number of molecules  \\
\hline
1			& PLac & 0 \\
2			& RNAP  & 40 \\ 
3			& PLacRNAP  & 1 \\
4			& TrLacZ1   & 0 \\
5			& TrLacZ2 & 15 \\ 
6			& TrLacY1 & 0 \\ 
7			& TrLacY2   & 1 \\ 
8			& RbsLacZ & 0 \\ 
9			& RbsLacY    & 1 \\ 
10			& Ribosome &471 \\ 
11			& RbsRibosomeLacZ & 38 \\ 
12			& RbsRibosomeLacY &35\\ 
13			& TrRbsLacZ &883 \\ 
14			& TrRbsLacY &332 \\ 
15			& LacZ &1880 \\ 
16			& LacY &1608 \\ 
17			& dgrLacZ &15 \\  
18			& dgrLacY &12 \\ 
19			& dgrRbsLacZ &37 \\ 
20			& dgrRbsLacY &24\\ 
21			& lactose &141918 \\ 
22			& LacZlactose & 48\\
23			& product		& 1673873\\
\hline
\end{tabular}
\end{table*}
\begin{table*}%[tbhp]
\centering
\caption{Partition for the LacZ/LacY reaction.}
\label{table:partition_LacZY}
\begin{tabular}{ll}
subset			& elements\\
\hline
$S_1$			& \{1, 5, 8, 13, 14, 17\}  \\
$S_2$			& \{2, 6, 10, 15, 16\} \\
$S_3$			& \{3, 7, 9, 12, 20\} \\
$S_4$			& \{4, 11, 21, 22\}  \\
$S_5$			& \{18, 19\}  \\
\hline
\end{tabular}
\end{table*}
\begin{table*}%[tbhp]
\centering
\caption{Initial population and diffusion rate of a large-scale reaction-diffusion system.} \label{table:rd}
\begin{tabular}{llcc}
		id			& reacting species  & number of molecules & diffusion rate \\
\hline
1			& p53 & 38.3355 & 0.5648\\
2			& Mdm2  & 1.81548 & 0.3727\\ 
3			& Mdm2$\_$mRNA  & 38.1101 & 0.3465\\
4			& p53$\_$Mdm2   & 15.0388 & 0.9453 \\
5			& ATM$\_$I & 0 &0.4528 \\ 
6			& ATM$\_$A & 0 & 0.0801\\ 
7			& p53P   & 0.000191227 & 0.1599\\ 
8			& p300 & 8.99129 & 0.007136\\ 
9			& p53p$\_$p300    & 671.386 & 0.0483 \\ 
10			& p53$\_$A  &18465.5 & 0.6786\\ 
11			& HDAC1 &  2.66895 & 5.491\\ 
12			& Mdm2$\_$SMAR1$\_$HDAC1 &0.000243163 & 0.0795\\  
13			& p53$\_$Mdm2$\_$p300 &675.173 & 0.1973\\ 
14			& Mdm2$\_$p300 &0.257587 &0.7926 \\ 
15			& SMAR1 &0.00516598 & 0.08357\\ 
16			& Mdm2$\_$SMAR1 &0.170117 & 0.6910\\ 
17			& p53$\_$Mdm2$\_$SMAR1 &0.0777557 & 0.1425\\  
18			& Mdm2$\_$HDAC1 &0 & 0.1439 \\
\hline
\end{tabular}
\end{table*}

\begin{table*}[tbhp]
\centering
\caption{Thermochemical coefficients of the species in the H2-air mixture reaction (part I).}
\begin{tabular}{lccccc}
species & $a_1$ & $a_2$ & $a_3$ & $a_4$ & $a_5$ \\
\hline
H		& 0.000000000	& 0.000000000 & 2.500000000 & 0.000000000 & 0.000000000  \\ 
  		& 6.078774250e1	&-1.819354417e1 & 2.500211817 &-1.226512864e-7 & 3.732876330e-11 \\
O		&-7.953611300e3	& 1.607177787e2 & 1.966226438 & 1.013670310e-3 &-1.110415423e-6  \\
  		& 2.619020262e5	&-7.298722030e2 & 3.317177270 &-4.281334360e-4 & 1.036104594e-7  \\                      
H2O		&-3.947960830e4	& 5.755731020e2 & 9.317826530e-1 & 7.222712860e-3 &-7.342557370e-6  \\
  		& 1.034972096e6	&-2.412698562e3 & 4.646110780 & 2.291998307e-3 &-6.836830480e-7  \\            	                                                          
OH 		&-1.998858990e3 	& 9.300136160e1 & 3.050854229 & 1.529529288e-3 &-3.157890998e-6\\
  		& 1.017393379e6 	&-2.509957276e3 & 5.116547860 & 1.305299930e-4 &-8.284322260e-8  \\
O2		&-3.425563420e4 	& 4.847000970e2 & 1.119010961 & 4.293889240e-3 &-6.836300520e-7 \\
 		&-1.037939022e6 	& 2.344830282e3 & 1.819732036 & 1.267847582e-3 &-2.188067988e-7  \\
H2		& 4.078323210e4 	&-8.009186040e2 & 8.214702010 &-1.269714457e-2 & 1.753605076e-5 \\
 		& 5.608128010e5 	&-8.371504740e2 & 2.975364532 & 1.252249124e-3 &-3.740716190e-7\\
H2O2	    &-9.279533580e4 	& 1.564748385e3 &-5.976460140 & 3.270744520e-2 &-3.932193260e-5  \\
 		& 1.489428027e6 	&-5.170821780e3 & 1.128204970e1 &-8.042397790e-5 &-1.818383769e-8 \\  	                       
HO2 		&-7.598882540e4 	& 1.329383918e3 &-4.677388240 & 2.508308202e-2 &-3.006551588e-5  \\
 		&-1.810669724e6 	& 4.963192030e3 &-1.039498992 & 4.560148530e-3 &-1.061859447e-6 \\
N2		& 2.210371497e4 	&-3.818461820e2 & 6.08273836 &-8.530914410e-3 & 1.384646189e-5  \\  
 		& 5.877124060e5 	&-2.239249073e3 & 6.06694922 &-6.139685500e-4 & 1.491806679e-7 \\ 		
\hline
\end{tabular}
\label{thermodata}
\end{table*} 
\begin{table*}[tbhp]
\centering
\caption{Thermochemical coefficients of the species in the H2-air mixture reaction (part II).}
\begin{tabular}{lccccc}
species & $a_6$ & $a_7$ & $b_1$ & $b_2$ \\
\hline 		
H		& 0.000000000     & 0.000000000     & 2.547370801e4 &-4.466828530e-1 \\ 
  		&-5.687744560e-15 & 3.410210197e-19 & 2.547486398e+4 &-4.481917770e-1 \\
O		& 6.517507500e-10 &-1.584779251e-13 & 2.840362437e+4 & 8.404241820 \\
  		&-9.438304330e-12 & 2.725038297e-16 & 3.392428060e+4 &-6.679585350e-1 \\                      
H2O		& 4.955043490e-09 &-1.336933246e-12 &-3.303974310e+4 & 1.724205775e+1 \\
  		& 9.426468930e-11 &-4.822380530e-15 &-1.384286509e+4 &-7.978148510 \\            	                                                          
OH 		& 3.315446180e-9 &-1.138762683e-12 & 2.991214235e+3 & 4.674110790\\
  		& 2.006475941e-11 &-1.556993656e-15 & 2.019640206e+4 &-1.101282337e1 \\
O2		&-2.023372700e-9 & 1.039040018e-12 &-3.391454870e+3 & 1.849699470e1 \\
 		& 2.053719572e-11 &-8.193467050e-16 &-1.689010929e+4 & 1.738716506e1 \\
H2		&-1.202860270e-8 & 3.368093490e-12 & 2.682484665e+3 &-3.043788844e1 \\
 		& 5.936625200e-11 &-3.606994100e-15 & 5.339824410e+3 &-2.202774769 \\
H2O2	    & 2.509255235e-8 &-6.465045290e-12 &-2.494004728e+4 & 5.877174180e1 \\
 		& 6.947265590e-12 &-4.827831900e-16 & 1.418251038e+4 &-4.650855660e1 \\  	                       
HO2 		& 1.895600056e-8 &-4.828567390e-12 &-5.873350960e+3 & 5.193602140e1 \\
 		& 1.144567878e-10 &-4.763064160e-15 &-3.200817190e+4 & 4.066850920e1 \\
N2		&-9.625793620e-9 & 2.519705809e-12 & 7.108460860e+2 &-1.076003744e1 \\  
 		&-1.923105485e-11 & 1.061954386e-15 & 1.283210415e+4 &-1.586640027e1 \\			
\hline
\end{tabular}
\end{table*} 

\end{document}